\documentclass[fleqn,twoside]{article}
\usepackage{espcrc2,epsfig}

\usepackage{graphicx}
\usepackage[figuresright]{rotating}

\newcommand{\AmS}{{\protect\the\textfont2
  A\kern-.1667em\lower.5ex\hbox{M}\kern-.125emS}}

\title{\vspace{-3mm}
Searching for KvBLL calorons in $SU(3)$ lattice gauge field ensembles\\
\vspace{-15mm}\normalsize $^{\;}$\hfill HU-EP-03/65
\vspace{13mm}}

\begin{document}
\thispagestyle{empty}

\author{Christof Gattringer\address[R]{Institut f\"ur theoretische
        Physik, Universit\"at Regensburg, D-93040 Regensburg,
        Germany}\thanks{Supported by Austrian Academy of Sciences APART~654.},
        Ernst-Michael Ilgenfritz\address[B]{Humboldt-Universit\"at zu 
        Berlin, Institut f\"ur Physik, D-12489 Berlin, Germany},
        Boris V. Martemyanov\address[M]{Institute of Theoretical 
        and Experimental 
        Physics, RU-117259 Moscow, Russia},
        Michael M\"uller-Preussker\addressmark[B],
        Dirk Peschka\addressmark[B],
        Rainer Pullirsch\addressmark[R],
        Stefan Schaefer\addressmark[R] and 
        Andreas Sch\"afer\addressmark[R]\thanks{This contribution is 
based on parallel talks at LATTICE 2003 by Christof Gattringer and
Ernst-Michael Ilgenfritz. This work is supported by DFG and BMBF.}}

\begin{abstract}
We discuss Kraan - van Baal - Lee - Lu (KvBLL) 
solutions of the classical Yang-Mills
equations with temperature in the context of $SU(3)$ lattice gauge theory.
We present discretized lattice versions of KvBLL solutions and other dyonic 
structures, obtained by cooling in order to 
understand their variety and signature.
An analysis of the zero modes of the lattice Dirac operator for different 
fermionic boundary conditions gives clear evidence for a KvBLL-like background 
of finite $T$ lattice subensembles with $Q=\pm1$.
Using APE-smearing we are able to study the topological charge density $q(x)$ 
of the configurations and to corroborate this interpretation. 
\end{abstract}
\maketitle

\section{INTRODUCTION AND RETROSPECT TO $SU(2)$}

The discovery of new caloron solutions 
with non-trivial holonomy \cite{NEW-CALORONS}
has revived hopes for a semiclassical description of Yang-Mills theory
at $T \ne 0$, in particular of the deconfinement transition. A new aspect 
of these solutions is the interplay between topological charge and the  
local and asymptotic Polyakov loop. One new  feature, the separation 
of a $Q=1$ object (caloron) into $N_c$ dyonic ($D$) constituents, 
happens only in 
some part of the parameter space which seems to be statistically enhanced in 
the confinement phase. 

For a couple of years some of us have been searching \cite{IMMSV} 
for this type of semiclassical background in $SU(2)$ lattice Monte
Carlo configurations at finite $T$, mostly using cooling. 
Besides of opening  a zoo of possible lattice solutions, 
a remarkable outcome was that, near $T_c$, the confinement phase is 
distinguished from deconfinement by a finite yield of non-trivial caloron 
configurations with 2 dyonic constituents. Sometimes they are separated into
peaks of action, identifiable also by different localizations of the
fermionic zero mode for periodic/antiperiodic temporal b.c., 
but always monopole pairs can be seen inside, such that the Polyakov loop 
$P(x)$ changes from $+1$ to $-1$. Somewhat unexpectedly, also $D\bar{D}$ 
pairs have been obtained by cooling as metastable solutions, with action 
$S \approx S_{inst}$ and equal-sign Polyakov loop. This happens only in the 
confined phase where the $D\bar{D}$ pairs originate from caloron-anti-caloron 
structures. Recently we have repeated the $SU(2)$ cooling studies at lower 
temperature. We found now that all calorons consist of overlapping action 
lumps, yet the Polyakov loop exhibits their non-trivial caloron nature.

\section{SIGNATURE OF $SU(3)$ CALORONS}

Let us now come to $SU(3)$. We pa\-ra\-me\-tri\-ze as 
${\cal P}_{\infty} = 
\mathrm{diag}(\mathrm{e}^{2\pi i\mu_1},\mathrm{e}^{2\pi i\mu_2},
\mathrm{e}^{2\pi i\mu_3})$
the asymptotic holonomy,  
with $\mu_1 \leq \mu_2 \leq \mu_3 \leq \mu_4 = 1+\mu_1$ 
and $\mu_1+\mu_2+\mu_3 = 0$.  Let $\vec{y}_1$, $\vec{y}_2$ and $\vec{y}_3$ be
three $3D$ position vectors remote from each other.
Then a KvBLL \cite{NEW-CALORONS} 
caloron consists of 3 lumps carrying the instanton action split 
into fractions $\mu_2 - \mu_1$, $\mu_3 - \mu_2$ and $\mu_4 - \mu_3$,
concentrated near the $\vec{y}_i$.

\begin{figure}[t]
\vspace*{-1mm}
\hspace*{1.5mm}
\epsfig{file=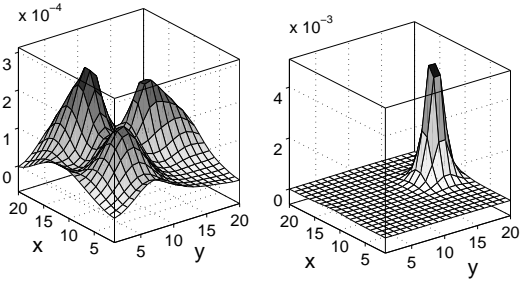,width=7cm,clip}
\vspace{-9mm}
\caption{Topological charge density in the~plane through all 3
constituent positions for a non\-trivial (l.h.s.) and a trivial (r.h.s) 
KvBLL~caloron.}
\vspace{-4mm}
\label{fig:nontrivial_vs_trivial}
\end{figure}

We discretized continuum calorons and applied a few cooling iterations
to adapt it to torus boundary condition for the gauge field.
Then an action $S \approx 1.3 S_{inst}$ and a 
topological charge $Q=\pm1$ are
found.
In Fig.~\ref{fig:nontrivial_vs_trivial} we show two extreme cases of lattice 
KvBLL solutions with well separated positions $\vec{y}_i$: a non-trivial one 
with almost equidistant phases $\mu_i$ and a caloron of trivial holonomy, 
{\it i.e.}, ${\cal P}_{\infty} \approx \mathrm{diag}(1,1,1)$, 
$\mu_1,\mu_2,\mu_3 \approx 0$. 
In the trivial case one of the constituents
%% , at $\vec{y}_3$, 
carries (almost) all action and topological charge.  

In Ref.~\cite{KVB-ZEROMODES} the zero mode $\psi_0$ of the Dirac operator in 
the background of a KvBLL solution was analytically computed, introducing 
an arbitrary phase in the temporal b.c.\ of the Dirac operator, {\it i.e.}, 
$\psi_0(t+1/T,\vec{x}) = \exp(2\pi i\zeta) \, \psi_0(t,\vec{x})$.   
An intricate interplay between the phase parameter $\zeta$ and the holonomy 
phases $\mu_i$ was found. The zero mode is not localized 
{\it simultaneously} at {\it all} dyonic constituents.   
Instead, it chooses {\it one} of them and changes its position with
changing value of $\zeta$. 
A selection rule states that the zero mode is localized at 
the position $\vec{y}_i$ of the $i$-th constituent if 
$\mu_i \leq \zeta \leq \mu_{i+1}$. This implies that 
for a set of well-separated $\mu_i$ {\it and} $\vec{y}_i$ 
the zero mode will visit all 3 constituents during one 
cycle of $\zeta$ from 0 to 1.~\footnote{For cooled $SU(2)$ configurations 
evidence for KvBLL-type configurations was 
obtained from periodic and anti-periodic 
zero modes of the Wilson-Dirac operator \cite{IMMSV}.  } 

\begin{figure}[t]
\epsfig{file=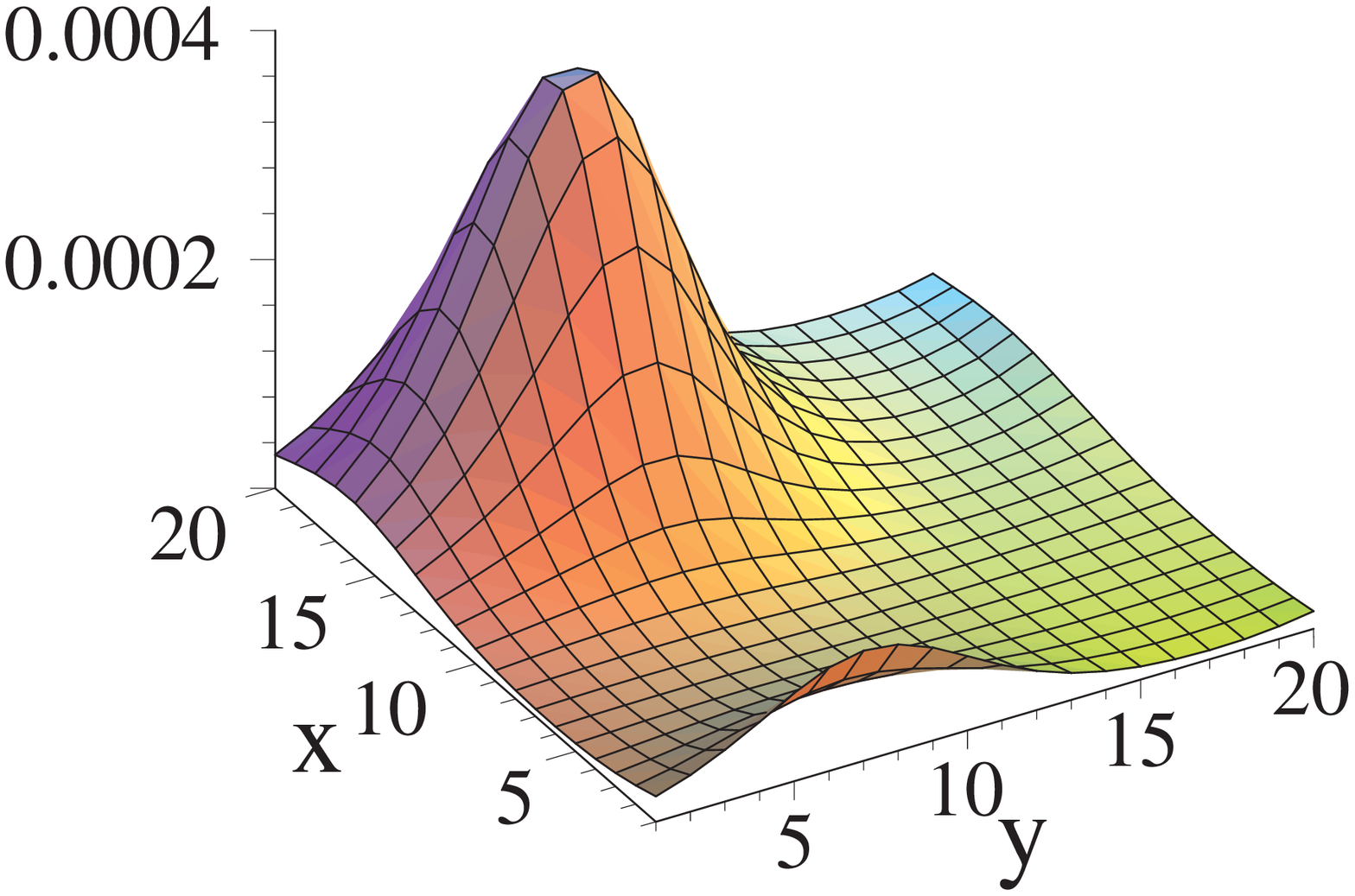,width=3.95cm}
\hspace{-6mm}
\epsfig{file=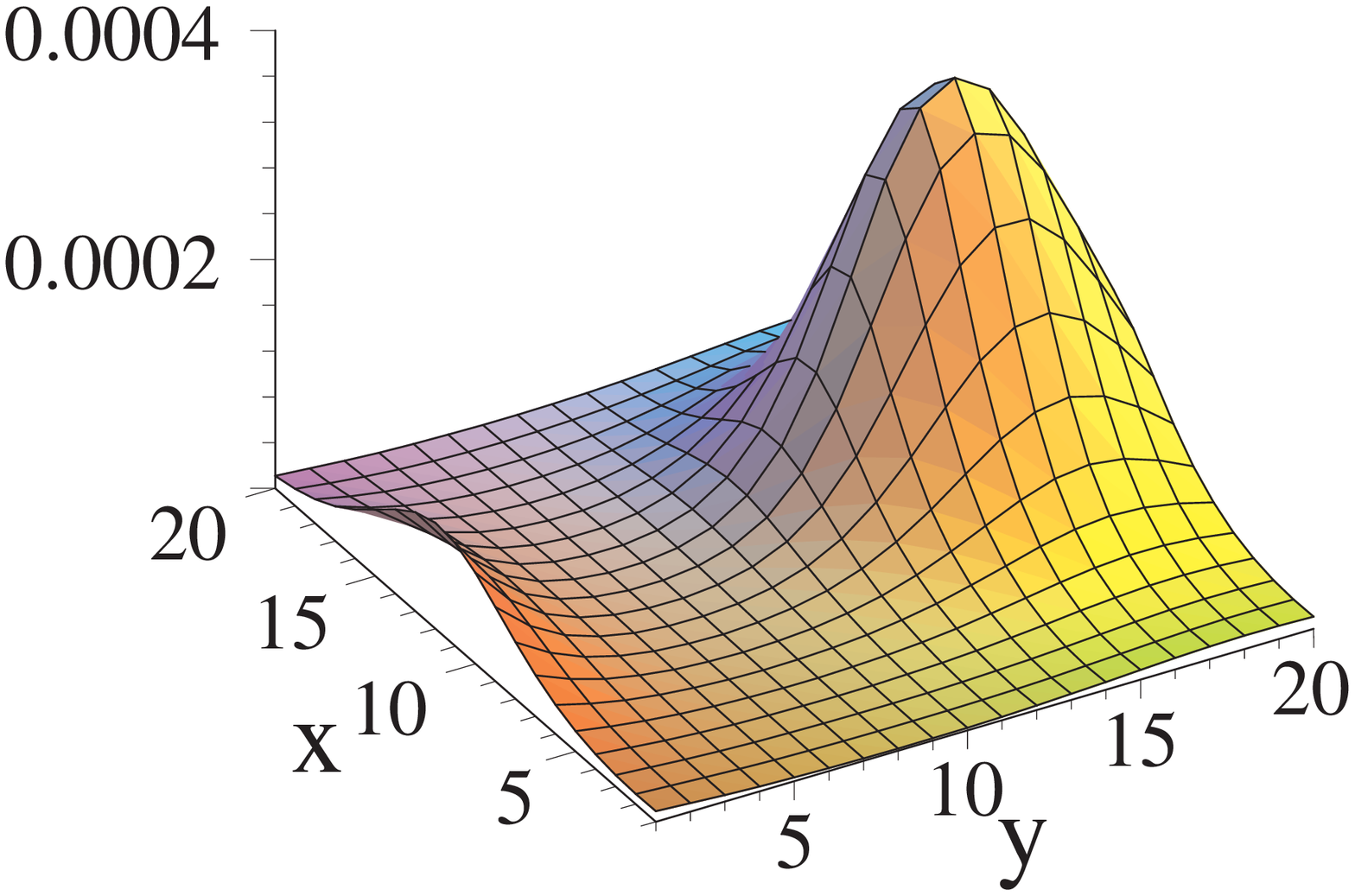,width=3.95cm}
\vspace{-9mm}
\caption{Scalar density for the non-trivial KvBLL caloron of Fig.\ 1. We show
2 zero modes obtained with $\zeta = 0.5$
(l.h.s.) and
$\zeta = 0.8$ (r.h.s.).}
\label{fig:rotating_zero_modes}
\vspace{-4mm}
\end{figure}

We demonstrate this effect using 
the chirally improved lattice Dirac operator \cite{CHIRAL-IMPROVED}.
This operator has good chiral properties and
is known to reproduce the zero mode of latticized standard instantons 
\cite{INSTANTON-TEST}. We clearly reproduce the behavior of the zero 
mode for the non-trivial caloron (see Fig.~\ref{fig:rotating_zero_modes}). 
We plot the scalar density of the zero mode. This 
gauge invariant observable is obtained by summing $|\psi_0(x)|^2$
at each lattice point $x$ over color and Dirac indices. 
In order to assess the physical relevance of calorons,
one should study the correlation between the zero mode positions 
and the topological charge density $q(x)$ and find out the
statistical distribution in the caloron parameter space. 
This could be difficult
because generic equilibrium configurations might contain dyonic constituents 
difficult to separate. Yet it is indispensable because the calorons have to 
be disentangled from the additional topological structure which might be 
of different nature. 

\section{COOLED CALORONS AND OTHER OBJECTS}
\begin{figure}[t]
\hspace{1mm}
\epsfig{file=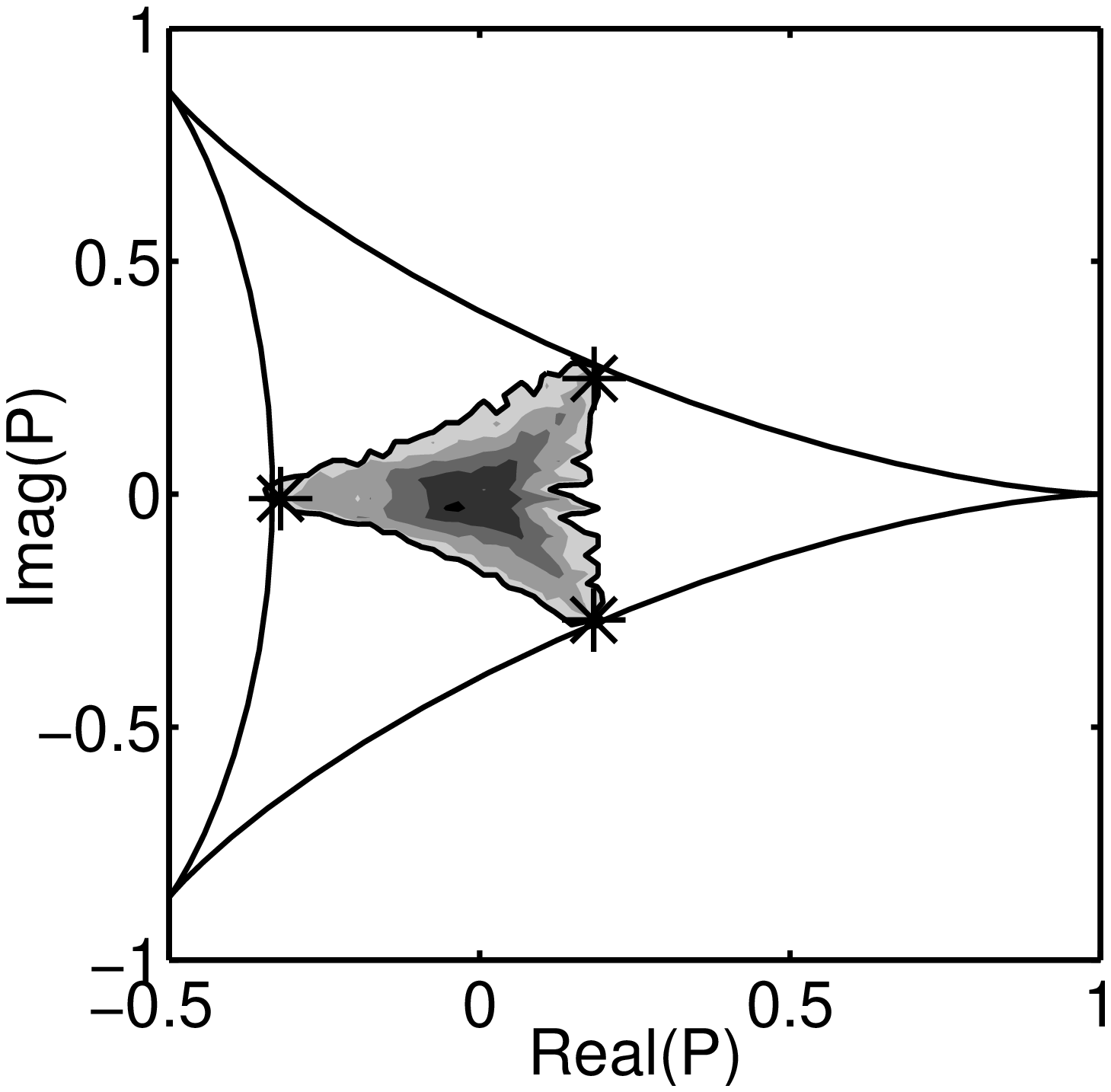,width=3.50cm}
\epsfig{file=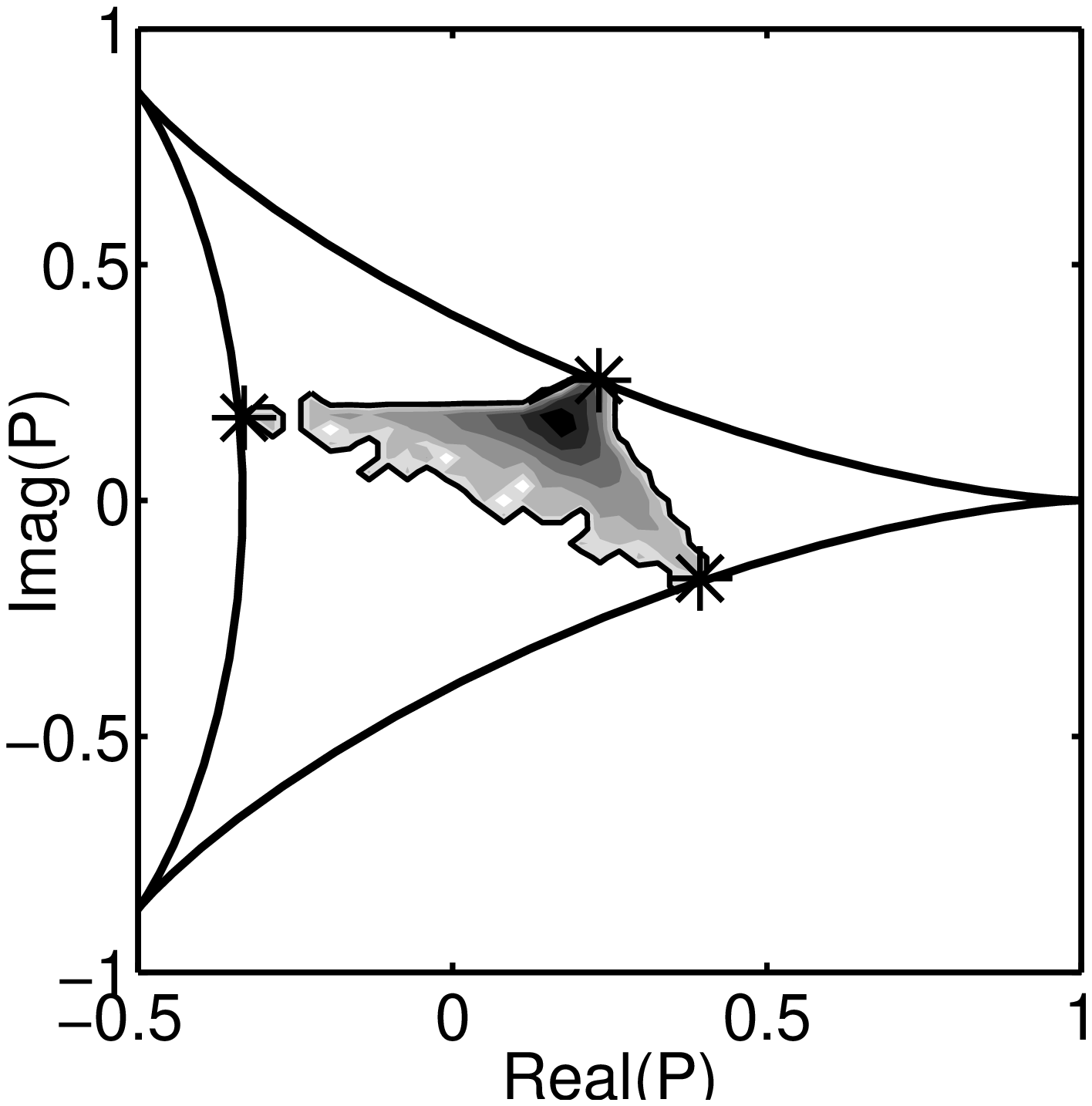,width=3.40cm}
\vspace{-7mm}
\caption{Polyakov loop distributions over the volume 
for the constructed non-trivial caloron of Fig. 1 
(l.h.s) and for a caloron obtained by cooling (r.h.s.).
Constituents are marked with stars.}
\label{fig:PL-histogram}
\vspace{-4mm}
\end{figure}

In order to explore the appearance of more general $SU(3)$ KvBLL solutions 
on the lattice we cooled hot configurations 
in the confinement phase on $16^3\times4$ lattices 
(created with Wilson gauge action at $\beta=5.5$), 
keeping torus b.c.\ for the gauge field. 
Then the evolution of the holonomy,
starting from an average Polyakov loop $\bar{P} \approx 0$
(where $P=\frac{1}{3}\mbox{tr}{\cal P}$), is not predictable. 
With increasing number of cooling steps $n_{cool}$ 
the asymptotic Polyakov loop $\bar{P}_{\infty}$ 
(an average over all lattice points 
where the local action density $s < 10^{-4}$) 
becomes more and more scattered 
over the complex region accessible for the Polyakov loop.
One can find cooled configurations with any non-trivial holonomy.
Quasiclassical lattice configurations are identified when the cooling
history $S(n_{cool})$ goes through a plateau
{\it i.e.}, the equations of motion are minimally violated.
The distribution in $\bar{P}_{\infty}$ is more centered around 
the origin for higher plateaus and more spread for the caloron 
plateaus $S=S_{inst}$. 
Of course, only part of the caloron events with asymptotic 
holonomy $\bar{P}_{\infty} \notin Z(3)$ have such a clean $DDD$ structure
of the action density as the non-trivial example shown 
in Fig.~\ref{fig:nontrivial_vs_trivial}. 
Most of the $S=S_{inst}$ plateaus have only two distinguishable 
lumps of action. Nevertheless, all caloron events are characterized 
by a distribution of the Polyakov loop field 
$P(\vec{x})$ as shown in Fig.~\ref{fig:PL-histogram} where the ideal 
non-trivial caloron is compared with a generic caloron (two-lump) event. 
The bulk of lattice points represents the asymptotic holonomy while the 
constituents are mapped to the envelope of this plot. In the metric of $P$, 
the distance of a constituent from the bulk is proportional to the fraction 
of action (topological charge) it carries.

Purely (anti-)selfdual configurations are most frequently observed on the 
$S = S_{inst}$ and $S= 2 S_{inst}$ plateaus. The latter, analogous to the 
calorons, preferentially contain four lumps of action, but probably belong 
to a wider class of $Q = \pm 2$ KvBLL solutions.
Besides of these, we have found metastable cooled configurations which 
are {\it not} of KvBLL caloron type. In the $Q=0$ sector we found $D\bar{D}$ 
events at plateaus $S \approx 0.65 S_{inst}$ as well as $DD\bar{D}\bar{D}$ 
events at $S \approx 1.35 S_{inst}$, usually left over from the annihilation 
of $D$'s and $\bar{D}$'s, former constituents of caloron and anti-caloron
in a confinement configuration.

\section{ZERO MODES FOR $Q = \pm 1$ EQUILIBRIUM CONFIGURATIONS}

In the last few years it was understood that the low lying eigenmodes
of the lattice Dirac operator provide a good filter which removes short 
distance fluctuations of the gauge field allowing the study of topological
objects on the lattice. This mostly concerns the {\it near zero modes}.
The peculiar 
%% phase dependence 
behavior of the {\it zero mode} known for the KvBLL 
configurations suggests that this $\zeta$ dependence for {\it any} 
zero mode might be an excellent, simple and stable tool to highlight the 
presence of caloron background fields in equilibrium configurations 
among other carriers of topological charge, without cooling.
A more detailed account of the results presented in this section can be 
found in Ref.~\cite{REGENSBURG}. 

We have computed eigenmodes of the chirally improved lattice Dirac 
operator \cite{CHIRAL-IMPROVED}, using quenched $SU(3)$ gauge 
ensembles generated with the L\"uscher-Weisz action 
\cite{LUESCHER-WEISZ-ACTION} 
on $20^3 \times 6$ lattices at inverse gauge couplings of $\beta = 8.20$ 
and $\beta = 8.45$. These values correspond to lattice spacings of 
$a = 0.115$ fm and $a = 0.094$ fm and 
%% give rise to ensembles 
describe physics just below and above the QCD phase transition 
(located at $\beta_c = 8.24$~\cite{GAUGE-DATA}). 
From the full ensembles of gauge configurations we have selected those 
configurations which have only a single zero mode, {\it i.e.}, configurations 
with topological charge $Q = \pm 1$. This choice leads one to consider only 
the simplest case where no mixing between different zero modes can occur. 
All or a certain fraction of these configurations might contain the 
excess topological charge $\pm 1$ in the form of well-separated dyons.

\begin{figure}[t]
\epsfig{file=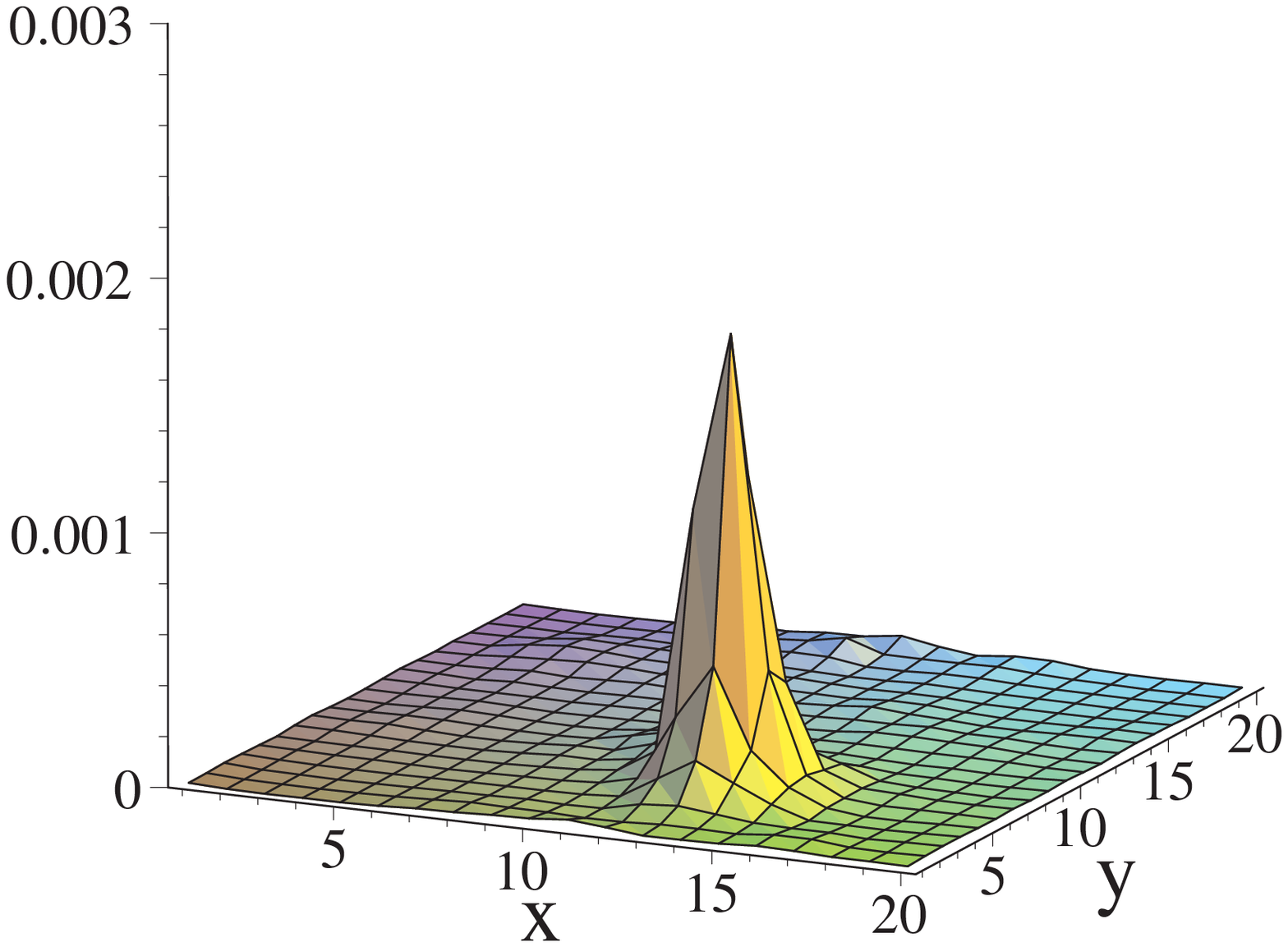,width=3.95cm}
\hspace{-6mm}
\epsfig{file=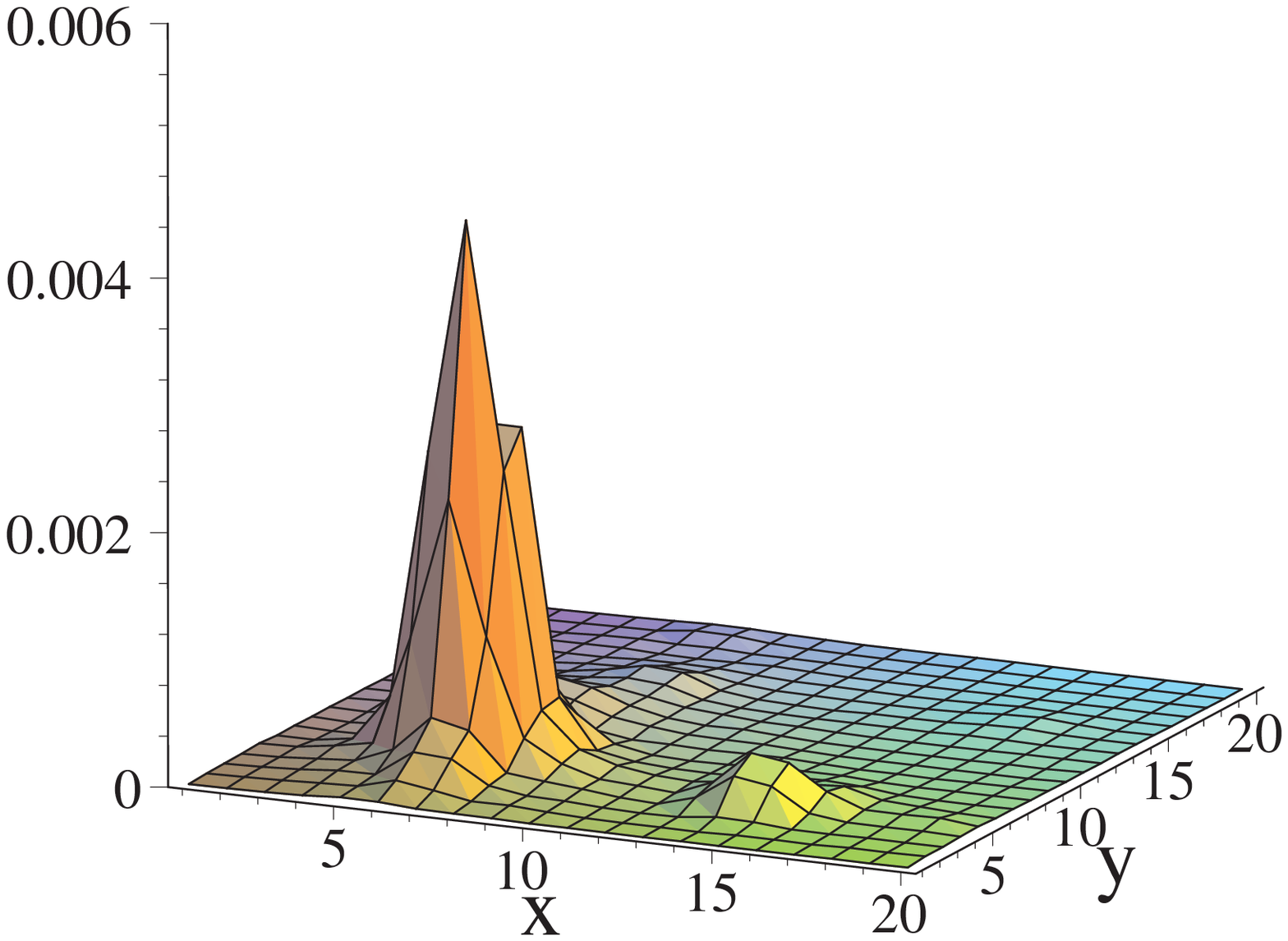,width=3.95cm}
\vspace{-9mm}
\caption{Scalar density of the zero mode of config.\ No.\ 125
(confined phase) for anti-periodic (l.h.s.) and periodic b.c. (r.h.s.).
In each case we show the $x$-$y$ slice containing the maximum.}
\label{fig:rhoplot}
\vspace{-5mm}
\end{figure}

For an equilibrium configuration, let us illustrate the effect of changing 
fermionic b.c. on the scalar density of the zero mode $\psi_0$. 
In Fig.~\ref{fig:rhoplot} we show $x$-$y$ slices of the scalar density 
of the zero mode for the configuration No.\ 125 from the 
confined ensemble and compare the cases of anti-periodic and periodic temporal 
fermionic boundary conditions. The position of the zero mode is seen to 
change under a switch from $\zeta=0$ to $\zeta=0.5$ which gives us 2 out of 
3 expected locations. 
Later we will show that in all 3 positions the zero mode 
is located on top of lumps of topological charge with appropriate sign.
In the particular example shown in Fig.~\ref{fig:rhoplot} the two positions 
are 1.3 fm apart.  

For a more systematical study 
we computed the zero mode 10 times on the same configuration using values 
of $\zeta = 0.0, 0.1, 0.2, ... , 0.9$ in the fermionic b.c.. 
This detailed investigation was done for 10 configurations in the 
confined phase ($\beta = 8.20$) and 10 configurations in the deconfined 
phase ($\beta = 8.45$). 

In the confined phase the expectation value of the Polyakov loop
vanishes, suggesting a choice 
$\mu_1 = -1/3$, 
$\mu_2 = 0$, 
$\mu_3 = 1/3$, $
\mu_4 = 2/3$. 
The selection rule then implies that $\zeta$ can be 
contained in three different intervals $[0,1/3]$, $[1/3,2/3]$, $[2/3,1]$, 
each selecting another dyonic constituent. For 5 of our 10 configurations 
we found that the zero mode indeed visits exactly 3 different lumps. 
For 3 of the 10 configurations we found only 2 distinct lumps. 
The explanation might be that two of the constituents are very close to each 
other in space. Finally, for 2 of the 10 configurations we found even 4 
different positions of the zero mode. This latter observation could be an 
effect of large quantum fluctuations on 
top of the infrared structure mimicking 
an additional peak. It would be interesting 
to repeat the calculation on a finer 
lattice with the same temperature where one 
expects the disturbance from quantum 
fluctuations to be less severe.

In the deconfined phase the Polyakov loop has a non-vanishing expectation
value. Due to the center symmetry of the gauge action this expectation
value can be realized with three different $Z(3)$ phases $0, \pm 2 \pi /3$. 
For real Polyakov loop one has 
$\mu_1 \approx \mu_2 \approx \mu_3 \approx 0$ and 
$\mu_4 \approx 1$. Roughly speaking, 
the boundary phase parameter $\zeta$ can be 
contained only in the two intervals $[0,0]$ and $[0,1]$. 
Only for $\zeta \approx 0$ 
the zero mode is not able to see the same lump that it sees for all other 
$\zeta \neq 0$, which contains a well-localized topological charge 
($Q \approx \pm 1$). For configurations with complex Polyakov loop 
(phase $\pm 2 \pi /3$) this critical value of $\zeta$ is shifted to $1/3$ 
respectively $2/3$. Our set of 10 configurations 
in the deconfined phase contained 
configurations with real Polyakov loop as well as with complex Polyakov loop 
(here we also computed the values $\zeta = 1/3, 2/3$). For all 10
configurations we confirmed that only for the critical value of $\zeta$ 
a different position of the zero mode may be taken. 

\begin{figure}[t]
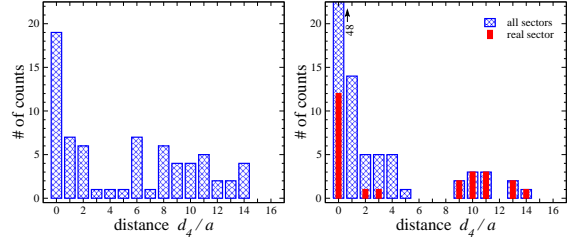

%% \begin{center}
\epsfig{file=disthisto6x20b820.eps, width=3.65cm, clip}
\epsfig{file=disthisto6x20b845.eps, width=3.65cm, clip}
\vspace{-7mm}
\caption{Histograms of the distance between the peak positions of 
the periodic and the anti-periodic zero mode in the confined (l.h.s)
and deconfined phase (r.h.s.).}
\vspace{-4mm}
\label{fig:disthisto}
\end{figure}

For a larger set of 89 deconfined configurations and 70 confined ones 
we followed the numerically 
cheaper approach of comparing only the periodic ($\zeta = 0.0$) with the
the anti-periodic ($\zeta = 0.5$) zero mode. For both boundary conditions
we determined the position of the corresponding peak and then computed the 
Euclidean distance $d_4$ between the two peaks. The histograms of distances 
are shown in Fig.~\ref{fig:disthisto}.
In the confinement phase (l.h.s.) $d_4$ is distributed over all distances. 
The histogram for deconfinement (r.h.s.) shows that for many configurations 
the mode does not jump at all or only over a small $d_4$. There is only a
small, separate component in the histogram with $d_4 \geq 9a$, all related 
to configurations having a real Polyakov loop $\bar{P}$. 
We highlight this subsample of the deconfined phase in the r.h.s. plot.
This observation confirms the selection 
rule for zero modes in a KvBLL background. 
It allows jumps between periodic and anti-periodic b.c. only in the real 
sector. This does, however, not imply that the zero mode in the real
sector {\it always} jumps over large distances 
since the two constituents might 
be close to each other. In Ref.~\cite{REGENSBURG} several other tests of the 
zero modes were performed such as a study of their inverse participation
ratio, a measure of their localization, as function of $\zeta$. 
Throughout we found fair agreement with the properties of zero modes in 
the background of a KvBLL solution 
%% with a holonomy consistent with the 
in the confined phase and even excellent agreement in the case of the 
deconfined phase. 
We have also extended our study of zero modes with general boundary conditions
to lattices with zero temperature geometry and still find jumping of the 
zero modes. A detailed writeup of these observations is in preparation.

In the meantime KvBLL solutions have been generalized to higher topological
charge, and their corresponding zero modes have been constructed analytically 
\cite{HIGHER-CHARGE}. It would be interesting to perform a study of
the zero modes in higher topological sectors. We also remark that in 
Ref.~\cite{INSTANTON-OVERLAP} it was argued that the effect of jumping zero
modes can hardly be due to overlapping standard instantons and therefore
this effect is truly a consequence of the presence of a KvBLL background 
in the $Q = \pm 1$ configurations under study.

\section{COMPARISON WITH SMOOTHING}

We have experimented with cooling techniques in order to study the correlation
between the topological charge density and the positions 
occupied by the zero mode with changing $\zeta$. 
We tried restricted cooling \cite{RIC}, adapted  to the L\"uscher-Weisz action,
and APE smearing. The latter procedure was used by A. Hasenfratz 
{\it et al.}~\cite{HDG} 
to disclose the topological content of lattice fields. 
In the confinement phase the emerging structures are similar for both
procedures (which therefore can be tuned to each other) and relatively 
stable with respect to cooling/smearing iterations. 

\begin{figure}[t]
\epsfig{file=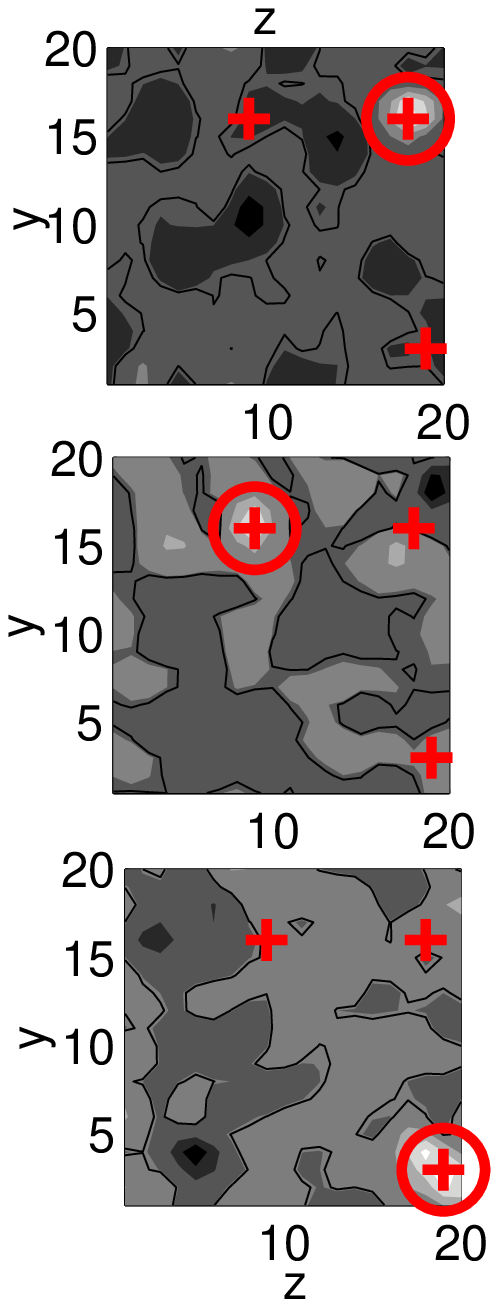,width=3.95cm}
\hspace{-6mm}
\epsfig{file=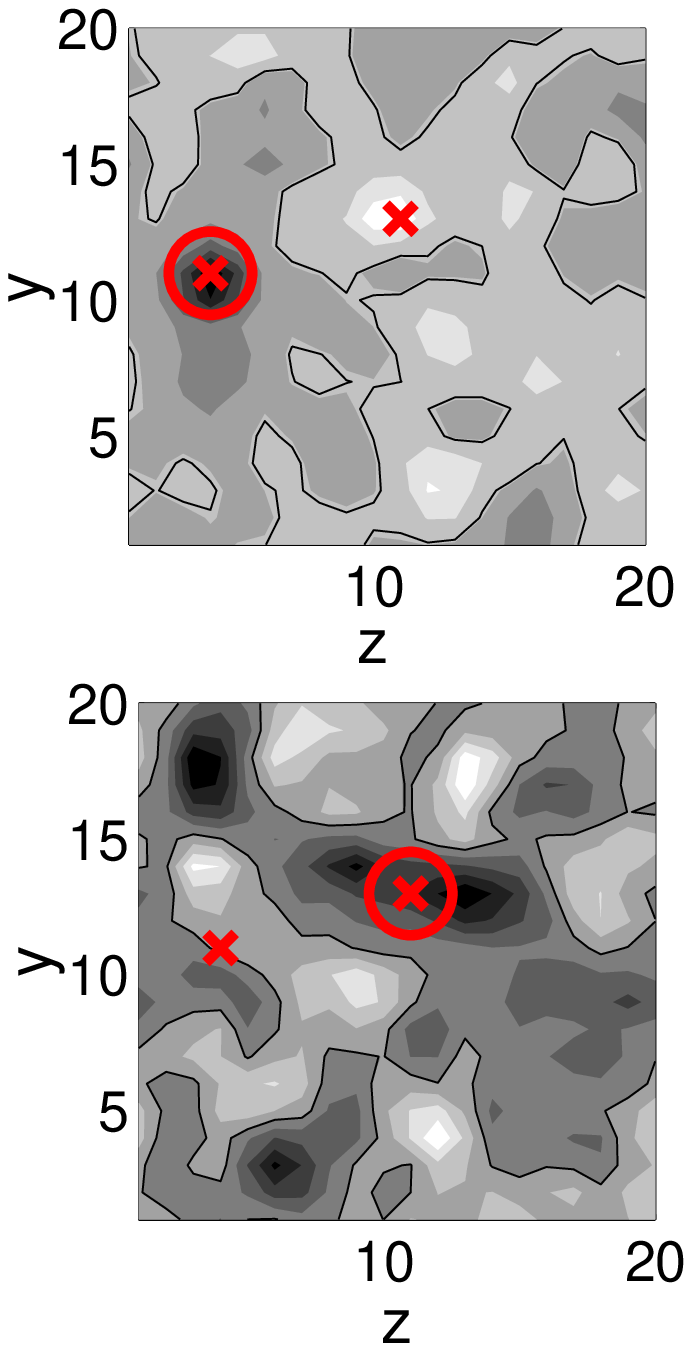,width=3.95cm}
\vspace{-9mm}
\caption{Topological charge density in $y$-$z$ planes, each through one
of the zero mode positions (marked by the circle): 
confined config. No.\ 125 with 3 different localizations (l.h.s.)
and deconfined config. No.\ 383 with 2 different localizations (r.h.s.).} 
\label{fig:topdensity}
\vspace{-6mm}
\end{figure}

Here we concentrate on results obtained by APE smearing  
with $N_{smear}=10$ iterations and $\alpha=0.45$ \cite{HDG}. 
As an example for the confined phase we show configuration No.\ 125
on the l.h.s. of Fig.~\ref{fig:topdensity}. 
The map of the topological density is 
shown in three $y$-$z$ slices, selected such that each contains one of the 3 
zero mode positions. 
Crosses show the projection of all 3 positions on the given
plane, while {\it crosses with circles} emphasize the zero mode sitting 
in the plane.  
There is a complex pattern of positive and negative charge 
density\footnote{The line separates regions of positive and negative $q(x)$.}
while the total charge is $Q = -1$. 
The zero mode always settles at same-sign lumps which are all well-localized. 
The topological charge is not exactly balanced between the lumps, because 
the largest $\zeta$ interval corresponds to 
$t_{zm}=5$, $x_{zm}=14$, $y_{zm}=16$, $z_{zm}=18$ 
(the position in the upper plot).  

As an example for the deconfined phase we show configuration No.\ 383 
on the r.h.s. of Fig.~\ref{fig:topdensity}. 
The configuration has $Q = 1$ and a 
real-valued asymptotic Polyakov loop. For almost all $\zeta$ the zero mode is 
localized on top of the well-localized constituent seen in the upper plot. 
For $\zeta \approx 0$ however, the zero mode appears much less-localized in 
another region of topological density $q(x) > 0$.
The topological charge is much less concentrated there, too,
consisting of two lumps with the zero mode center in between.

Obviously, also in the deconfined phase one 
finds a complex topological structure
which is more sensitive to cooling than in the confinement phase.
The gap separating the near zero modes strongly depends on it.
%% the distribution of topological charge. Hence, we 
We have observed that the gap rapidly closes if we cool too much.

\begin{figure}[t]
\vspace*{-2mm}
\hspace*{5mm}
\epsfig{file=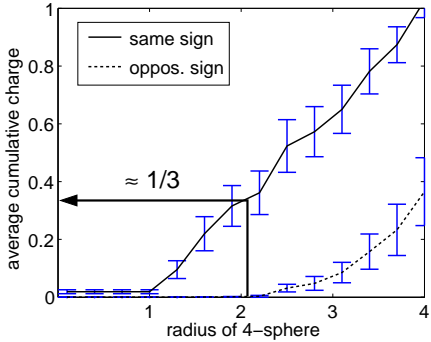,width=6.30cm}
\vspace{-8mm}
\caption{Cumulative topological charge of same sign and opposite sign
as function of radius $R$ around the center of a zero mode 
(in the confined config. No.\ 125).}
\label{fig:cumulative}
\vspace{-5mm}
\end{figure}

For the confined configuration No.\ 125 and for the three possible 
localizations of the zero mode we calculated
the cumulative topological charges of same and opposite sign 
(relative to $Q$) as a function of the radius of a 4-sphere. 
The curves shown in Fig.~\ref{fig:cumulative}
are an average over the three cases.
The radius inside of which the topological density has a coherent sign 
of $q(x)$ is a good definition of the constituent radius. 
Then the average topological charge of a constituent is 
$Q_{const} \approx \pm \frac{1}{3}$ in the confinement phase.

\section{CONCLUSIONS}

We obtained new classical solutions by cooling of confined $SU(3)$ 
lattice configurations.
For Monte Carlo configurations in the topological sector $Q = \pm 1$, 
the dependence of the localization of the fermionic zero mode on the 
aperiodicity angle $\zeta$ matches the specific expectations derived 
from the KvBLL caloron picture for the confined and deconfined phases. 
For slightly APE-smeared configurations we are able to corroborate 
these findings by considering the topological charge distribution 
close to the zero mode locations.

\end{document}